\documentclass[%
 reprint,
 amsmath,amssymb,
 aps,
]{revtex4-2}

\usepackage{graphicx}
\usepackage{dcolumn}
\usepackage{bm}

\begin{document}

\preprint{APS/123-QED}

\title{Plasmon localization and giant fields in
holographic metasurface for SERS sensors}

\author{Andrey K. Sarychev}
\email{sarychev${_}$andrey@yahoo.com}
\author{Andrey Ivanov}
\author{Andrey N. Lagarkov}
\author{Ilya Ryzhikov}
\author{Konstantin Afanasev}
\affiliation{%
	Institute for Theoretical and Applied 
	Electrodynamics, Russian Academy of Sciences, 125412, 
	Moscow, Russia}
\author{Igor Bykov}
\affiliation{%
	Institute for Theoretical and Applied 
	Electrodynamics, Russian Academy of Sciences, 125412, 
	Moscow, Russia; \\
	Laboratory of
	Nano-Bioengineering, 
	National Research Nuclear University MEPhI (Moscow
	Engineering Physics Institute), 
	31 Kashirskoe shosse, Moscow 115521, Russia
}%

\author{Gr\'egory Barbillon}

\affiliation{
EPF-Ecole d'Ing\'enieurs, 
3 bis rue Lakanal, 92330 
Sceaux, France\\
}%

\author{Nikita Bakholdin}
\author{Mikhail Mikhailov}
\affiliation{%
National Research University “Moscow Power 
Engineering Institute”, Krasnokazarmennaya 14, Moscow 
111250, Russia
}%

\author{Alexander Smyk}
\author{Alexander Shurygin}
\affiliation{%
	James River Branch llc, 8, Tvardovsky st., 
	Strogino Industrial Park Moscow 123458, Russia
}%

\author{Alexander 
	Shalygin}
\affiliation{%
	Lomonosov Moscow State University, Faculty of 
	Physics, GSP-1, 1-2 Leninskiye Gory, 	Moscow 119991, 
	Russia
}%

\date{\today}
\begin{abstract}
We present SERS-active metal 
holographic metasurfaces fabricated from metal
periodical nanograting deposited 
on a dielectric substrate. 
The metasurface consists of a modulated 
dielectric, which is covered 
by a thin silver layer. The metasurface 
operates as an open plasmon resonator. 
The theory of plasmons 
excited in the open resonator 
formed by a metal nanograting is presented. 
The large local electromagnetic
field is predicted for optical frequencies. 
The excitation of plasmons is 
experimentally demonstrated in the metasurface 
designed on a 4-inch Si wafer.
The enhancement of the  local 
electric field results in
surface-enhanced Raman scattering (SERS). 
To investigate the SERS effect, 
the metasurfaces are covered by molecules 
of 4-mercaptophenylboronic acid, 
which form covalent bonds with 
the silver nanolayer and serve as a proof-of-concept. 
Finally, we obtain a detection limit of 
230 nM for molecules of 4-mercaptophenylboronic acid. 

\end{abstract}

\maketitle


\section{Introduction}
The design of a nanoresonator with a high 
enhancement of the local electric field is an important 
point for the application to the surface-enhanced Raman 
scattering (SERS) and other plasmon-enhanced 
spectroscopies. It is important to increase the 
selectivity and sensitivity of the biological and 
chemical sensing 
\cite{Sharma2012,Yan2012,LizM2016,GB402019}. 
The novel fabrication techniques allow to produce 
sophisticated resonant substrates with the 
controllable shape and arrangement of the resonators 
\cite{BarbillonIvanovSarychev2020}. These 
techniques include focused ion-beam lithography 
\cite{Henzie2009}, electron-beam lithography 
\cite{Yu2008,GB2008,Manfrinato2019,GB2021}, 
X-ray, UV, plasmonic cavity lens (PCL) and 
interference lithographs
\cite{ZhangP2014,GB062008,VoDinh2011,%
	Quilis2018,Hwang2018,Gong2019}, 
nanoimprint Lithography (NIL) 
\cite{DingT2014,Farcau2019,Goetz2020}, nanosphere 
lithography (NSL) 
\cite{Masson2010,Bechelany2010,GB2018,%
	ChauY2019}, 
laser-induced transfer metal nanodroplets 
\cite{KuznetsovKiyanChichkov2010}. Some of them 
allow to realize reproducible SERS substrates with a 
lower cost and a larger active surface area, for 
example, interference lithography. In references
\cite{ZhangP2014,Hwang2018}, the gold and hybrid 
gold-silver nanodisk array fabricated via X-ray 
interference lithography is proposed. The 
effectiveness of the structures has been proven by the 
detection of a low concentration of Rhodamine 6G 
(R6G) molecules via SERS with enhancement factor 
$10^{5}-10^{6}$. The authors admitted high 
sensitivity, reproducibility and stability of such 
structures. In reference 
\cite{Gong2019}, silver nanohole arrays 
are proposed. A photoresist layer sandwiched with 
two Ag layers increases Q-factor of 
nanoresonators and the sensitivity. The SERS 
substrates exhibited enhancement factors (EFs) of 
$10^{7}$ that are capable of monolayer detection of 
R6G molecules. It should be noted that 
optically mirror underlayer, which is designed to 
increase Q-factor, was theoretically predicted and 
experimentally investigated by the authors 
\cite{GB402019,BarbillonIvanovSarychev2019,%
	LagarkovBoginskayaBykov2017,ZhouQ2010}. 
Theoretical predictions 
and experimental studies have shown giant 
electromagnetic field fluctuations in the case of 
almost touched plasmonic nanoparticles 
\cite{Genov2004,SealK2005,IvanovA2012,%
	Frumin2013,LiuZ2013,LiuG2014,Liu2014,%
	Rasskazov2013,SarychevIvanovBarbillon2021QE}. 
It was shown that plasmonic nanocavities confining 
the light to unprecedentedly small volumes, support 
multiple types of modes. Different nature of these 
modes leads to the mode beating 
within the nanocavity and the Rabi oscillations, 
which alter the spatio-temporal dynamics of the 
hybrid system \cite{DemetriadouA2017}. By 
intermixing plasmonic excitation in 
nanoparticle arrays with excitons in a $WS_2$ 
monolayer inside a resonant metal microcavity, the 
hierarchical system was fabricated with the collective 
microcavity-plasmon-exciton Rabi splitting exceeding 
$0.5\, eV$ 
at room temperature. Gap-surface plasmon 
metasurfaces, which consist of a subwavelength thin 
dielectric spacer sandwiched between an optically 
thick film of metal and arrays of metal 
subwavelength elements arranged in a strictly or 
quasiperiodic fashion, 
have a possibility to fully control the amplitude,
phase, and polarization of the reflected light
\cite{IvanovA2012,DingF2018}.

Open resonators can very effectively accumulate the 
electromagnetic energy (see, e.g., 
\cite{Vainshtein1969}). In the optical frequency 
range, a corrugated metal surface 
can operate as an open resonator and generate large 
local electric field 
\cite{DykhneSarychev2003,DengMoskovits2010,%
	LagarkovSar2016,KanipMoskovits2016}. 
The method of the transformation optics was developed
to calculate optical properties of metamaterials 
\cite{PendrySchurigSmith2006,Leonhardt2006}. 
This method was used to investigate reflection
from the metal gratings in recent papers
\cite{KraftLuoMaier2015,HuidobroChangKraft2017,%
	PendryHuidobroLuo2017,PendryHuidobroDing2019}. 
The authors develop an
original conformal mapping that transforms
one-side metal grating into rectangular metal slab that
optical properties can be found. For instance, 
to calculate the reflection of an electromagnetic
wave, the metal grating is replaced by thin 
plane metal film. The developed method was used
to theoretically investigate broadband THz absorption in
graphene metasurfaces
\cite{GaliffiPendryHuidobro2018},
optical properties of singular metal surfaces
\cite{PendryHuidobroLuo2017,YangHuidobroPendry2018}, 
non-local effects \cite{YangGaliffiHuidobro2020},
and calculating of the energy loss of an electron flying
over the metal modulation \cite{YangHuidobroPendry2020}. 

Strong motivation for the investigation of open 
plasmon resonators is the SERS effect, which is 
mainly due to the local electric 
field enhancement. The SERS effect 
enables identification of trace molecules captured 
by the corrugated metal surface. 
The SERS is extremely important for 
medical diagnostics, for instance, for cancer 
detection, imaging and therapy, drug delivery, 
quantitative control of biomarkers including glycated 
proteins and cardiovascular biomarkers 
\cite{DongL2017,Kneipp2017,HuY2017,
	AndreouC2016,Nechaeva2020,ChonH2014}. 

In this paper, we present the analytical theory 
as well as experimental observation of the 
plasmons excited in open resonators formed by a 
periodic metal grating. 
The resonance condition of the open plasmon
resonator is found, 
when it is possible to achieve large electromagnetic 
field enhancement both inside and in the vicinity of 
the plasmonic grating. 
This will increase the sensitivity of
SERS-probing and other surface-enhanced 
spectroscopies. 
The modulated silver film
is produced by four-ray holographic process 
and its optical properties are measured. 
The silver  metasurfaces are covered 
by molecules of 4-mercaptophenylboronic acid 
to investigate the SERS 
in metasurfaces made by optical interference
lithography. Holographic 
SERS substrates are easy-made and low-cost for 
mass production, and they have allowed to achieve a 
detection limit of 230 nM for molecules of 
4-mercaptophenylboronic acid (4-mPBA). 
\section{Analytical theory} 
We consider periodic metal gratings that support 
localized plasmons whose size can reach 
few hundred nanometers. The metal gratings with 
two-side modulation efficiently concentrate and 
store electromagnetic energy. The local electric field 
is resonantly enhanced. The localized plasmons are 
analytically calculated using a quasistatic 
approximation for the nanogratings whose 
dimensions (thickness and modulation) are much 
smaller than the wavelength $\lambda$. 
\subsection{Plasmon resonance in metal
	nanotube}\label{ring}
To understand how the metasurface works, let us 
solve a simple two-dimensional plasmonic system, 
which is interesting by itself since 
it has a non-trivial analytical 
solution. 
We investigate the plasmon resonance and the
corresponding electromagnetic field
enhancement in the metal nanotube that radius 
and thickness are much smaller than the 
wavelength $ \lambda $. 
The length of the metal nanotube is much larger
than the radius. The vector of the external 
electric field lies in the $ u - v $ plane
perpendicular to tube axis. 
Since the transversal dimensions of 
the nanotube is much smaller than the wavelength
the problem is reduced to the two dimensional
problem of the excitation of the infinite metal
nanotube by the $ E = \{E_{u}, E_{v}\} $ electric 
field, which is normal to the tube axis.
For further simplification, we neglect retardation 
effects and use the quasistatic approximation 
when the plasmonic electric field is found from the 
electric potential $ \varphi(u, v) $. 
The two dimensional problem can be solved
by considering potential,  electric field, and charge  in 
$ u - v $
plane using theory of functions of a complex
variable. All the distances are measured in terms 
of the internal radius of the nanotube. 

\begin{figure}[ht]
	\centering{\includegraphics[scale=0.2]{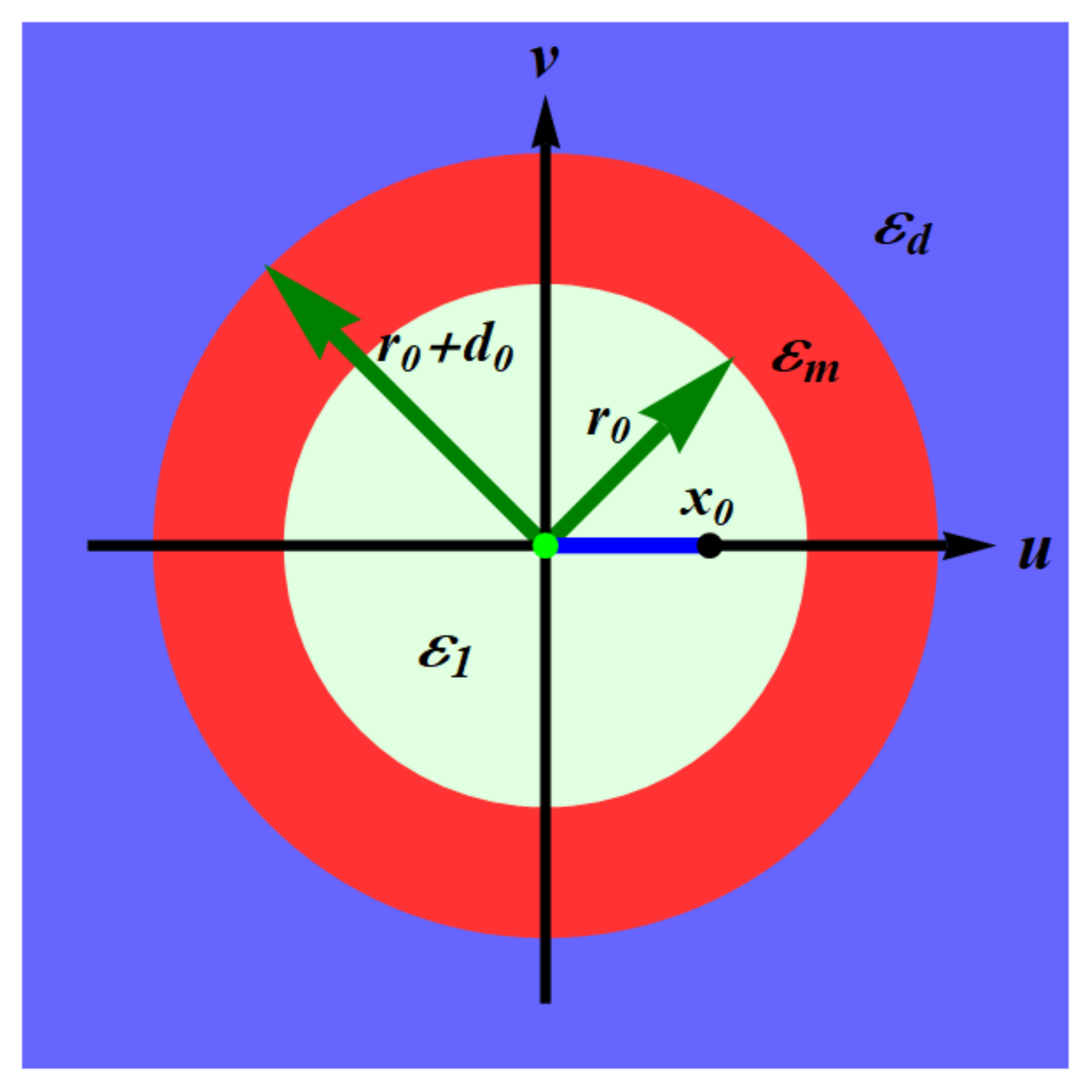}}
	\caption{Metal disk in plane $ u-v $ with 
		permittivity of internal space 
		$ \varepsilon_1 $, metal permittivity 
		$ \varepsilon_m $, 
		permittivity of outer space 
		$ \varepsilon_d $, and a "magnetic" monopole 
		placed at $ u = x_0, v = 0 $. }
	\label{Fig1}
\end{figure}

Thus, we calculate the plasmon
resonance and the field enhancement in a metal ring with radius 
$ r_{0}$, the thickness $d_{0}$, 
and the metal permittivity $\varepsilon_{m}$, 
shown in Fig.\,\ref{Fig1} by red color. 
To find the field in the ring, we introduce the 
Descartes coordinates $u$, $v$, and the complex 
variable $w = u + i v$. The ring has the center at 
$u = v = 0$. It is surrounded by a dielectric host 
$\varepsilon_{d}$, blue color in Fig.\,\ref{Fig1};
the medium inside the ring has permittivity 
$\varepsilon_{1}$ (light green in Fig.\,\ref{Fig1}). 
The electromagnetic field is 
excited by the point "magnetic" monopole, 
which generates in the surrounding space the electric 
potential $ \varphi_{s} = \Re \varphi_{sing} $, where
the complex potential equals to
\begin{equation}\label{Fsing}
	\varphi_{sing} = i  E_{0} \log ( w - x_{0}),
\end{equation}
where  $u = x_{0}$ and $v = 0$  are 
coordinates of the monopole  
(horizontal blue line in Fig.\,1). 
The exact nature of the
monopole is irrelevant for the material presented below.

\noindent
The monopole oscillates with the frequency 
$ \omega $, and the amplitude $ E_{0} $ is assumed,
for simplicity, to be positive $ E_{0} >0 $. 
Recall the quasistatic approximation is used, 
when the field retardation is neglected. Then, the 
electric potential $\varphi$ is the solution of the 
Laplace equation 
$ \triangle \varphi = 0 $
that holds in all the space but the monopole center.
As usual in two-dimensional (2D) problem, 
a complex electric field is introduced 
$ \tilde{E} = E_{u} - i E_{v} = - d \varphi / d w $. 
The field is expanded into radial 
$ E_{r} = \Re (w \tilde{E} )/ r$ 
$\equiv ( E_{u} u + E_{v} v ) / r$
and angular 
$ E_{\phi} = -\Im ( w \tilde{E} )/ r $
$ \equiv (- E_{u} v + E_{v} u ) / r $
components, where
$ r = \left| w \right| =\sqrt{u^{2} + v^{2}} $.
The electric charges, induced by the monopole, are 
uniformly distributed in the metal ring 
when the  monopole 
is placed in the center. Therefore, 
angular component of the electric field remains
to be the same inside and outside the ring. 
For the central monopole $ (x_{0} = 0) $ with amplitude  
$ E_{0} >0 $, the interior electric field 
$ E^{(i)} (r < 1 ) $, 
the field 
$ E^{(m)} ( 1 \le r \le 1+d_{0} )$
in the ring, and the outer field 
$ E^{(e)} ( r > 1+d_{0} ) $
have only angular components. 
The interior, ring, and outer fields are all equal to 
$ E^{(i)}_{r} = E^{(m)}_{r} = E^{(e)}_{r} = 0 $
and $ E^{(i)}_{\varphi} = E^{(m)}_{\varphi} =
E^{(e)}_{\varphi} = E_{0}/r $. 
As we see, the central monopole does not excite the 
plasmon regardless of the value of the metal 
permittivity $\varepsilon_{m}$. Yet, this is 
an unstable solution. Plasmons are excited when 
monopole is shifted from the center. 
Suppose that the 
monopole is set at $u = x_{0} <1, v = 0$. The 
interior complex potential equals to
\begin{eqnarray}\label{Fw} 
	\varphi^{(i)}(w) = 	
	\varphi_{sing}(w) +\varphi^{(i)}_{reg}(w) =\nonumber\\
	i E_{0} [\log( w - x_{0}) 
	+ \varphi^{(i)}_{reg}(w) ], \;\;  r \le 1
\end{eqnarray}

where 
$\varphi^{(i)}_{reg}(w)$ is a regular part of the 
interior potential that can be expressed in series of 
$w$ that only contains positive powers. To simplify
the consideration, we neglect for a moment 
the loss in the metal assuming that 
permittivity $ \varepsilon_{m} $ 
has zero imaginary part. 
The complex interior field equals to
\begin{eqnarray}\label{Ei1}
&	\tilde{E}^{(i)}(w) =- \dfrac{d\,\varphi (w)}{d\,w}
	= 	
	\\  \nonumber
&	- i E_{0} \left[ \frac{ 1 }{ w - x_{0}} 
	+  \sum _{ n = 1 }^{\infty } A_n w^{n-1} x_0^n 
	\right], \;\; r \le 1,
\end{eqnarray}
where
$ 0 < x_{0} <1 $
and it is still assumed that $ E_{0} $ is real. 
The first term in 
Eq.\,(\ref{Ei1}) is
expanded in series of 
$ (x_{0}/w )^{n}$
for 
$ x_{0} < \left| w \right| <1 $, and we obtain
\begin{eqnarray}\label{Ei2}
& \tilde{E}^{(i)}(w) =
- i E_{0} \left[ w^{-1}+ \right.
\\ \nonumber
&
\left. \sum _{ n = 1 }^{\infty } x_0^n 
\left(A_n w^{n-1} + 
w^{-n-1}\right)\right] ,
\;\; r \le 1
\end{eqnarray}
where coefficients $A_{n}$ take real values. 
The electric field in the 
metal ring is expanded in Laurent series:
\begin{eqnarray} \label{Em}
&\tilde{E}^{(m)}(w) = - i E_{0} \left[ w^{-1} +\right.
\\ \nonumber
& \left. \sum _{n=1}^{\infty } x_0^n \left(B_n w^{-n-1}
	+
	C_n w^{n-1}\right)\right]
	, \;\; 1<r<1 + d
\end{eqnarray}
and the outer complex field decreases when the radius
$ r \equiv \left| w \right| $ goes to infinity:
\begin{eqnarray} \label{Ee}
& \tilde{E}^{(e)}(w) = - i E_{0} \left[ w^{-1}  + 
 \sum _{n=1}^{\infty } 
x_0^n D_n w^{-n-1} \right],
\\ \nonumber
&  r \ge 1 + d
\end{eqnarray}
These complex fields can be split in radial and 
angular parts, namely,
\[ 
E = \left\lbrace E_{r}, 
E_{\phi} \right\rbrace =
\left\lbrace 
\Re	\left[ \tilde{E} \frac{w}{r}\right] ,
- \Im \left[ \tilde{E} \frac{w}{r}\right] 
\right\rbrace. \]
For example, the internal field $ 
E^{(i)} $ has the following radial and angular 
components
\begin{widetext}
\begin{equation}\label{Ei3}
	\left\lbrace E^{(i)}_{r}, 
	E^{(i)}_{\phi} \right\rbrace =
	E_{0} \left\lbrace 0, r^{-1} \right\rbrace \\
	 + E_{0}
	\sum_{n=1}^{\infty} r^{-n-1} x_0^{n} \left\{ 
	\left(A_n r^{2 n}-1\right) \sin (n \phi), 
	\left(A_n r^{2 n}+1\right) \cos (n\phi) \right\}, 
	\; r \le 1 
\end{equation}
\end{widetext}

The $ 	\left\lbrace E_{r}, E_{\phi} \right\rbrace $
expansion of the internal $E^{(i)} $, 
ring $E^{(m)} $, and outer $E^{(e)} $ fields holds 
even if the coefficients $ A_{n}, 
B_{n}, C_{n}, D_{n} $ are complex values. 
The coefficients ascribe complex values since
the metal permittivity has imaginary part 
$ \varepsilon_{m} = 
\varepsilon_{m1} + i \varepsilon_{m2}$
due to ohmic loss. 

\begin{figure}[ht]
	\centering{\includegraphics[scale=0.3]{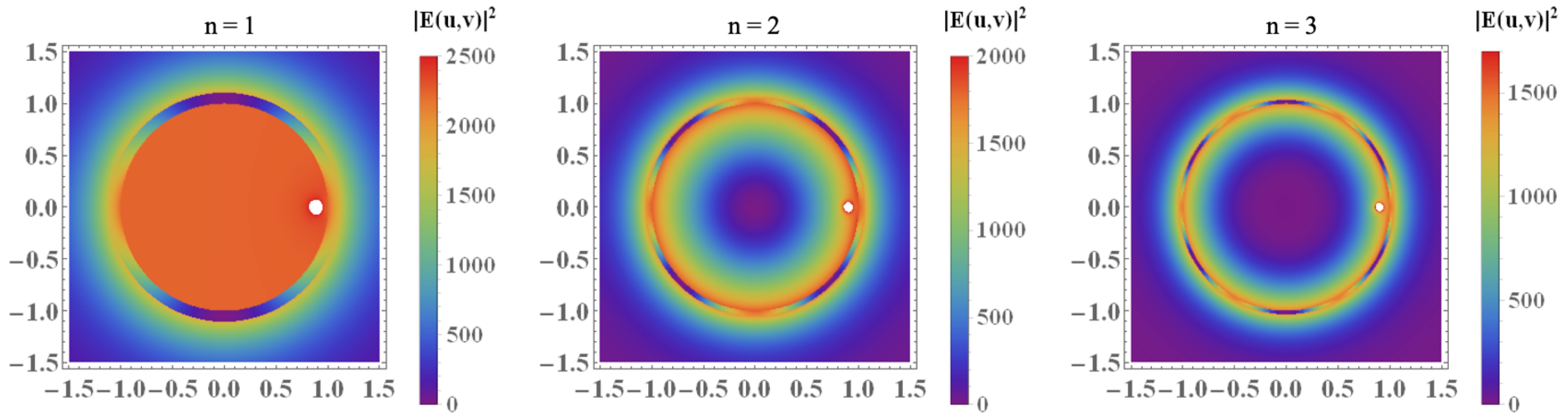}}
	\caption{
		Electric field in the silver ring for the 
		plasmon resonances $n = 1, 2, 3$ at an excitation 
		wavelength of $ \lambda = 785\,nm$. 
		Silver permittivity is 
		$\varepsilon_{m} \approx-30 + 0.4 i$, 
		internal permittivity 
		$\varepsilon_{1} = 1$, 
		outer permittivity $\varepsilon_{d} = 2$,
		ring thickness
		$d_{0res,n} = 0.11, \, 0.052, \, 0.034$, 
		all dimensions are normalized to 
		the ring radius $ r_{0} $. 
		The white spot corresponds to the
		monopole, which is placed at 
		$u = x_0 = 0.9, \, v = 0$. }
	\label{Fig2}
\end{figure}

The important value of the intensity 
$ I^{(i)}= \left| E^{(i)} \right|^{2}
/ \left| E_{0} \right|^{2}$ 
of the internal field, which is averaged over the 
internal surface of the metal ring, 
($ \left| w \right| \equiv r =1 $) equals to
\begin{widetext}
\begin{equation}\label{Iav} 
	\left\langle I^{(i)} \right\rangle =
	\dfrac{1}{ 2 \pi }\int_0^{2 \pi}
	\left| E^{(i)} (r = 1, \phi) \right| ^2 \, d\phi
	= 1 + \frac{1}{2} \sum_{n=1}^{\infty} 
	x_0^{2 n} \left( \left| A_n \right| ^2 + 1\right)
	\equiv 1 + \sum_{n=1}^{\infty} I_{n}
\end{equation}
\end{widetext}
where the coefficients $A_n$ 
are given below by Eq.\,(\ref{AnBnCnDn}). 
The coefficients $A_{n}, B_{n}, C_{n}, D_{n}$ in 
Eqs.\,(\ref{Ei1}) -- (\ref{Iav}) are found from the 
boundary conditions for the electric 
fields at internal 
surface $(\left| w \right| = 1)$ and outer surface 
$(\left| w \right| = 1 + d_{0})$ of the ring:
\begin{eqnarray} \label{InSur}
&	\varepsilon_{1} E^{(i)}_{r} = \varepsilon_{m} E^{(m)}_{r} , \;
E^{(i)}_{\varphi} =
E^{(m)}_{\varphi}, \; 	\; \left| w \right| = 1,
\\ 	\label{OutSur}
&	\varepsilon_{m} E^{(m)}_{r} =
\varepsilon_{d} E^{(e)}_{r} , \;
E^{(m)}_{\varphi} =
E^{(e)}_{\varphi}, \; 	\; \left| w \right| = 1+d_{0}
\end{eqnarray}
where $\varepsilon_1$, 
$\varepsilon_m$ and, $\varepsilon_{d}$ 
are the permittivities of the internal space $( r \le 1) $,
metal ring $ (1 < r < 1 + d_{0}) $ and the outer space
$ (r \ge 1 + d_{0}) $, 
respectively. We substitute the series
(\ref{Ei1}) -- (\ref{Ei3}) 
in the above boundary equations and 
equate the coefficients 
at the same power of $x_{0}$. 
It is easy to verify that the coefficients at $x_{0}^{n}$ 
in all series (\ref{Ei1})-(\ref{Ei3}) 
have the same angular dependence. 
Thus, the following basic 
equations are obtained:
\begin{eqnarray} \label{ABC}
&	A_n+1= B_n+C_n, \;\; 
	\varepsilon_1 \left(A_n-1\right) = 
	\varepsilon_m \left(C_n-B_n\right),
\nonumber	\\ 
&	B_n+C_n d_{2 n}=D_n,\;\; 
	 \varepsilon_m \left(B_n - 
	C_n d_{2 n}\right)=\varepsilon_d D_n 
\end{eqnarray}
where $d_{2 n} = (1 + d_{0})^{2 n}$. Solution of
these equations gives the coefficients
\begin{eqnarray}\label{AnBnCnDn}
&	A_n=
\frac{d_{2 n} \left(\varepsilon_1-\varepsilon_m\right) 
\left(\varepsilon_d+\varepsilon_m\right)
-\left(\varepsilon_m+\varepsilon_1 \right) 
\left(\varepsilon_d-\varepsilon_m\right)}
{det_{n} }, 
\\ \nonumber
& B_n=\frac{2 d_{2 n} \left(
\varepsilon_d+\varepsilon_m\right)}{det_{n} },\;
C_n=\frac{2 \left(\varepsilon_m-
\varepsilon_d\right)}{det_{n}}, \;
D_n=\frac{4 d_{2 n} \varepsilon_m}{det_{n} }
\end{eqnarray}
that completely determine the electric field inside, in, and out
of the plasmonic ring. 
The determinant $ det_{n} $ of equations Eq.\,(\ref{ABC}) 
that is the denominator in Eqs.\,(\ref{AnBnCnDn}) equals to
\begin{eqnarray}\label{DD}
&	det_n = d_{2 n} \left( \varepsilon_m + 
\varepsilon_1 \right)
\left( \varepsilon_m + \varepsilon_2 \right) +
\nonumber\\ 
&	\left( \varepsilon_1 - \varepsilon_m \right) 
\left( \varepsilon_m - \varepsilon_d \right)
\end{eqnarray}
Zeroing of the determinant $ det_{n} = 0 $ gives the 
condition for the $n$\textit{-th} resonance exciting 
in the plasmonic ring. For the given permittivities of 
the metal ring $\varepsilon_m$ and outer space 
$\varepsilon_d$, the ring resonates when its 
thickness takes the following values:
\begin{eqnarray}
& \label{dRes}
d_{0res,n} = r_{0} \, 
\Re \left[ 
\left(\frac{\left(\varepsilon_1 - \varepsilon_m\right) 
\left(\varepsilon_d-\varepsilon_m\right)}
{\left(\varepsilon_m + \varepsilon_1 \right) 
\left(\varepsilon_d + 
\varepsilon_m\right)}\right)^{\frac{1}{2 n }}
- 1 \right] 
\simeq 
\nonumber  \\
&  r_{0}  \left[ 
\left(\frac{\left(\varepsilon_1 - \varepsilon_{m1}\right) 
\left(\varepsilon_d-\varepsilon_{m1}\right)}
{\left(\varepsilon_{m1} + \varepsilon_1 \right) 
\left(\varepsilon_d + 
\varepsilon_{m1}\right)}\right)^{\frac{1}{2 n }}
- 1 \right] 
\end{eqnarray}
where we restore natural dimensions and
insert the radius $ r_{0} $ of the ring. 
$n = 1,2,3, \ldots$ is the resonance number, 
i.e., $ 2 n $ is the number of nodes in the ring 
field $ E^{(m)} $ (see Fig.\,\ref{Fig2}). 
In transition to the second estimation in Eq.\,(\ref{dRes}),
it is taken into account that the real part 
$ \varepsilon_{m1} = \Re [\varepsilon_{m}] $
of the metal permittivity is much larger in absolute value
than the imaginary part 
$ \varepsilon_{m2} = \Im [\varepsilon_{m}]
\ll \left| \varepsilon_{m1} \right| $
for the "good" optical metals like silver or gold. 
We obtain an expected 
result: the ring resonates if and only if the real part 
of the metal permittivity is negative. 
For the red and infrared parts of the optical spectrum, the 
real part 
$ \varepsilon_{m1} $
of the metal permittivity is negative and 
large in absolute value. Then, the dispersion 
equation Eq.\,(\ref{dRes}) simplifies to 
\begin{equation} \label{dRes1}
	d_{0res, n} \simeq r_{0} \frac{\varepsilon_d+\varepsilon_1}
	{n |\varepsilon_{m} |}	
\end{equation} 
where 
$ r_{0}$ is the radius of the ring. We obtain that 
the resonance thickness of the metal layer
is inverse proportional to the absolute value
of the metal permittivity. 

The dimensionless intensity 
$\left\langle I^{(i)} \right\rangle $ 
of the resonance electric field averaged over the internal 
surface of the ring with thickness $ d_{0res, n} $ 
estimates as 
$ \left\langle I^{(i)} (d = d_{0res, n} )\right\rangle =
I_{0 \max ,n} \simeq I_{n} $
from Eq.\,(\ref{Iav}):
\begin{eqnarray} \label{Imax} 
	I_{0 \max ,n} \simeq 
	2 \left[ 
	\frac{\epsilon_1 \epsilon _{m1}
		\left(\epsilon_d^2-\epsilon_{m1}^2\right)}
	{ \epsilon_{m2} 
		\left(\epsilon_d+\epsilon_1\right)
		\left(\epsilon_{m1}^2-\epsilon_1 \epsilon_d\right)}
	\right]^{2} 
	\left(\frac{x_0}{r_0}\right)^{2 n}
\end{eqnarray}
where $\varepsilon_{m1} = \Re \varepsilon_{m} $
and
$\varepsilon_{m2} = \Im \varepsilon_{m}$. 
Excitation of various resonances in the silver ring is 
illustrated in Fig.\,\ref{Fig2} for the wavelength 
$\lambda = 785 \, nm$. The silver permittivity for 
this wavelength is 
$\varepsilon_{m} \simeq -30 + 0.4 i$ 
\cite{Johnson1972}, and the permittivity of the 
outer space is chosen to be $\varepsilon_{d} =2$, 
which corresponds to the permittivity of the 
photoresist in our experiments. The first three 
resonances are displayed in Figure \ref{Fig2}. 
The electric field in a metal ring has two, four 
and six nodes for the first, second and 
third resonances, respectively. Note
the first dipole resonance electric field,
shown in Figure\,\ref{Fig2}a,
fills all the internal space of the ring. 
This result corresponds to the electric field distribution 
in the metal cylinder excited by an uniform external electric field
\cite{LandauLifshitzPitaevskii1984}. 
A plasmon excited in a metal tube can be 
experimentally investigated. Suppose that the
nanochannels with radius $ r_{0} $
are formed in a dielectric
material. Nanochannels can
be produced, for instance, in the silicon by
the electrochemical etching (see 
\cite{MariazziBettottiLarcheri2010} 
and reference therein). 
The surface of the nanochannels is 
covered by a silver layer with the
thickness 
$d_{res, 1} \simeq r_{0} (\varepsilon_d+1) /
( \left| \varepsilon_{m} \right| )$ 
given by Eq.\,(\ref{dRes1}). 
Suppose further that the Raman-active analyte is 
injected in these channels. Then, the Raman signal 
from the molecules
in the nanochannels is much enhanced since the 
scattering is incoherent and the Raman intensities 
are summed up. 
The Raman enhancement factor (SERS) estimates as 
$G \sim I^{2}_{max,n}$
(see, e.g., \cite{BrouersBlacherLagarkov1997}), where 
$I_{max,n} \gg 1$ 
is given by Eq.\,(\ref{Imax}) and 
the parameter $x_{0} \sim r_{0} $. 
For the typical Raman spectroscopy wavelength
of $ \lambda = 785\,nm $, 
the silver permittivity is about
$\varepsilon_{m} \simeq -30 + 0.4 i$
and the enhancement factor estimates as 
$ G \sim \left( \varepsilon _{m1}/
\varepsilon_{m2} \right) ^4 \sim 10^{7}$. 
This esteem holds for any argument of the complex
amplitude $ E_{0} $. For instance, in the case
$  E_{0} = - 2 i Q,  \, Q>0 $ 
the monopole potential (\ref{Fsing}) is the potential
of  the usual electric charge, which  density equals 
to $ Q \delta(u-x_{0})  \delta( v ) $. 
Any other field  sources such as a dipole
or  quadrupole, can be combined from 
electric monopoles. 
\subsection{Giant electric field in metal 
	grating}\label{grating}
\begin{figure}[ht]
	\centering{\includegraphics[scale=0.25]{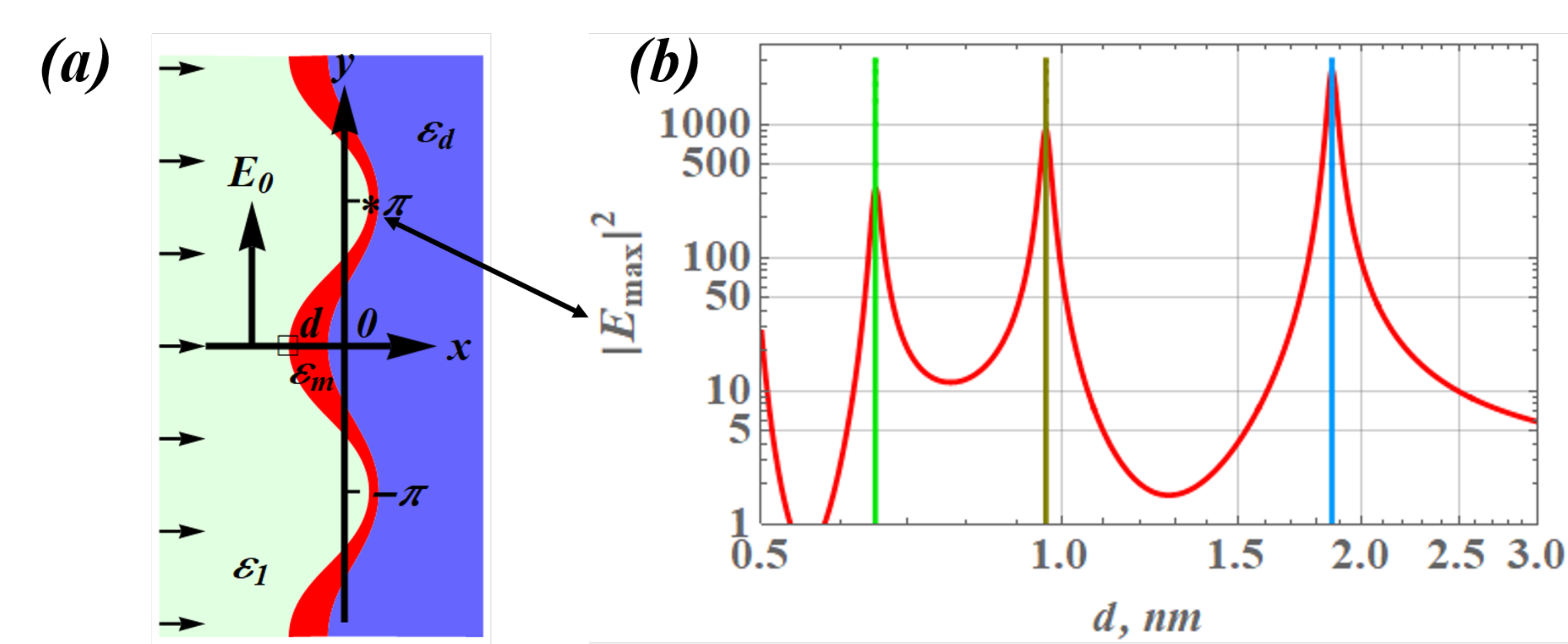}}
	\caption{\textit{\textbf{(a)}} 
		The design of plasmonic grating based 
		on a metal film with a permittivity 
		$\varepsilon_{m}$ deposited on the modulated 
		dielectric substrate (shown in blue) with a 
		permittivity $\varepsilon_{d}$,
		metal film has thickness $ d $, 
		"\textit{y}"-polarized light is incident from left;
		star "$ \ast $" denotes the point with coordinates 
		$ x = x^{(f)}_{max}$ and $y = \pi$, 
		where surface electric 
		field has maximum; small open square "$\square$"
		denotes the point with coordinates 
		$ x = x^{(f)}_{min}$ and $ y = 0 $, 
		where surface electric 
		field has minimum, the film modulation 
		$ h = \left(  x^{(f)}_{max} - 
		x^{(f)}_{min} \right)  (L/2 \pi) $. 
		\textit{\textbf{(b) }} Intensity of 
		local electric field at surface of silver film, which
		is modulated with period \textit{L}; the field is shown 
		in the point 
		$ X_{max} = x^{(f)}_{max} (L/2 \pi) $ and
		$ Y_{max} = \pi (L/2 \pi)$ 
		[star "$ \ast $" in \textit{\textbf{ (a)}}]
		as function of the 
		film thickness $d$, where the period is equal to 
		$L = 50 \,nm$, amplitude of the modulation is 
		equal to $h = 11 \,nm$, silver permittivity 
		equals to $\varepsilon_{m} \approx -30 + 0.4 i$ at 
		785 $nm$, permittivity of dielectric 
		substrate equals to $\varepsilon_{d} = 2$,
		permittivity of front medium $\varepsilon_{1} = 1 $. 
		Vertical color lines correspond to plasmon 
		resonances given by Eq.\,(\ref{dResXY}). }
	\label{Fig3}
\end{figure}

We consider a modulated metal film. The metal 
grating is placed at the plane $X = 0$. 
The electromagnetic wave propagates along 
{\it X}-axis being $ '' y '' $-polarized,
i.e., $E \sim \{0, E_0, 0 \} \exp(i k X)$, 
where $k = \omega/ c$ is the wave-vector. The 
wave impinges on the front surface of the grating, 
where the plasmon is excited due to the modulation
(see Figure \ref{Fig3}a). 
In this section devoted to the analytical 
consideration, we assume that the film modulation 
$h$ as well as 
its period $L$ are much smaller than the wavelength 
$ \lambda = 2 \pi/k $. In this case, we can 
approximate 
$E \sim \{0, E_0, 0 \} \exp(i k X) \simeq 
\{0, E_0, 0 \}$ 
that is the plasmon grating is excited 
by uniform field 
$E_{0}$, which is $y$-directed and oscillates with 
frequency $ \omega $. 
It is convenient to introduce the complex variable   
$ Z = X + i Y $. We use the quasistatic approach,
therefore, the incident field is described by 
the complex potential 
$\varphi (Z) = i Z E_{0},$ 
where we assume that the field 
$ E_{0} $ 
takes real values. Then, 
$E_{x} = -\Re[ d \varphi (Z)/ d Z ] =
0$, $E_{y} = \Im[ d \varphi (Z)/ d Z ] = E_{0}$ (see Fig.\ref{Fig3}). 
It is convenient to shift from the real coordinates 
\textit{X} and \textit{Y} to the dimensionless 
coordinates \textit{x} and \textit{y} that are defined 
by using the period of the film modulation $L$, 
namely, $x = 2 \pi X/L$ and $y = 2 \pi Y/L$. That is 
the periodically modulated metal film has
"natural" period of $2 \pi$ being considered in 
$x$ and $y$ frame as shown in 
Figure \ref{Fig3}a. 
Recall we use the  cuasistatic approximation, since 
the amplitude of metal modulation is assumed to be 
much smaller than the wavelength of the incident 
light. The potential of the local electric field does 
not change after re-scaling of the coordinates, since 
there is no characteristic length in the Laplace 
equation:
\begin{equation} \label{fz} 
	\varphi (z) = i z E_{0},
\end{equation}
where $ z = x + i y $.
The shape of the modulated metal is chosen 
in such way that the front surface 
$\{ x_{f}, y_{f} \}$ 
of the modulated metal film (left red surface in
Fig.\,\ref{Fig3}a) is given by 
the parametric equation, 
which it is convenient to write in a complex form
\begin{equation} \label{XY-UVf} 
	z^{( f )} ( \phi) = x^{(f)}( \phi) + i y^{{(f)}} 
	( \phi) = \log [\exp( i \phi) - x_0]
\end{equation}
where the parameter $x_0$ is in the 
interval $0 < x_0 < 1$, the variable $\phi$ changes 
in the limits $- \infty < \phi< \infty$; the logarithm 
is the full analytical function defined in all 
Riemann surfaces so that $y$ also varies within the 
limits of $- \infty < y < \infty$. The back surface 
$x_b ( \phi) > x_f ( \phi)$ of the metal film
(right red surface in Fig.\,\ref{Fig3}a)
is given by the parametric equation:
\begin{eqnarray} \label{XY-UVb}
	z^{(b)} ( \phi) =
		x^{(b)} ( \phi) + i y^{(b)} 
	( \phi ) = \nonumber\\
	\log [(1 + d_{0} )\exp( i \phi) - x_0]
\end{eqnarray} 
where the parameter $\phi$ still changes in the 
limits $- \infty < \phi < \infty$. The shape of the 
metal film is fully defined by its period $ L $
the amplitude of the modulation 
$ h = 
( x^{(f)}_{max} - x^{(f)}_{min}) \frac{L}{ 2 \pi}$ 
and the largest film thickness 
$d = (x^{(b)}_{min} - x^{(f)}_{min}) \frac{L}{ 2 \pi}$. 
The parameters $d_0$ and $x_0$ in the 
Eqs.\,(\ref{XY-UVf}) and (\ref{XY-UVb}) are 
expressed in terms of the film modulation $h$ and 
the film thickness $d$, namely,
\begin{equation}\label{xd}
	x_{0} = \tanh \left( \dfrac{h_{1}}{2} \right),\;\;
	d_{0} = 2 \dfrac{ \exp(d_{1}) -1 }{ \exp(h_{1})+1}
\end{equation}
where $ h_{1} = h (\frac{ 2 \pi }{ L } )$ and
$ d_{1} = d (\frac{ 2 \pi }{ L } )$. 
Therefore, the period $ L $, 
the modulation $h$, and thickness $d$ 
completely determine the 
shape of the modulated metal film indeed. 
Note that the film defined by 
Eqs.\,(\ref{XY-UVb}) and (\ref{XY-UVf})
mimics a metal film obtained, for example,
by the metal evaporation on the modulated
dielectric substrate. 
During the deposition process, the metal 
concentrates on tops of the film
(see Fig.\,\ref{Fig3}a and \ref{Fig7}). 
To find the optical electric field in and around 
a modulated metal film, illuminated by the 
incident light, we use new coordinates $u$ and $v$ 
that convenient to introduce in the complex form: 
\begin{equation}\label{zw}
	w = u + i v = \exp(z) + x_0, 
\end{equation}
where $z = x + i y$. In the transition 
from $x, y $ frame to $ u, v$ frame, the whole metal film 
transforms to the ring 
$1 \le \left| w \right| \le 1 + d_0$. 
The front and back surfaces of the grating are 
transformed to the internal $\left| w \right| = 1$ 
and outer $\left| w \right| = 1 + d_{0}$ surface of 
the ring. All the space in front of the grating (left 
from the grating in Fig.\,\ref{Fig3}a) transforms 
in the space inside the ring 
$\left| w \right| < 1$  in Fig.\,\ref{Fig1}a, 
the space behind the grating 
(right from the grating in Fig.\,\ref{Fig3}a) 
transforms into outer space of the ring 
$\left| w \right| > 1 + d_{0}$.  That is the metal 
film in Fig.\,\ref{Fig3}a is rolled into
the metal tube in Fig.\,\ref{Fig1}.
The approach is similar to the map 
used in references
\cite{IvanovA2012,KraftLuoMaier2015}. 
The electric potential given by Eq.\,(\ref{fz})
transforms in $\{u, v\}$ 
frame to the potential $ \varphi_{sing} $
in Eq.\,(\ref{Fsing}). Therefore, the 
complex electric potential $\varphi (z)$ 
for a modulated metal film equals to 
$\varphi (w(z))$, where $\varphi (w)$ 
is the potential, which is found in the previous 
subsection for the metal ring. 
The complex electric field inside and 
around the modulated 
metal film is obtained from the following 
equation 
$\tilde{E} (z) = E_{x} - i E_{y} = 
- d \varphi (w(z))/dz =
[- d \varphi (w)/dw] [dw/dz] = [dw/dz] 
\tilde{E} (w)$, where the complex field 
$\tilde{E} (w)$ is given by 
Eqs. (\ref{Ei1}) -- (\ref{Ei3}) and (\ref{AnBnCnDn}). 
The equation for the 
electric field ${E_{x} ,E_{y}}$ can be rewritten in 
the matrix form as follows:
\begin{equation} \label{MM}
\begin{bmatrix}
	E_{x} (x, y)\\
	E_{y} (x, y)
\end{bmatrix} =
\left[
\begin{array}{cc}
r -\frac{ u {x_0}}{r}   &   \frac{v {x_0}}{r}  \\
-\frac{v {x_0}}{r}      & r - \frac{ u {x_0}}{r}  \\
\end{array}
\right]
\begin{bmatrix}
E_{r} (u,v) \\
E_{\phi } (u,v) 
\end{bmatrix},
\end{equation}
where $\{E_{x},E_{y} \}$ is the field in the metal 
film (see Figure \ref{Fig3}\,a), $\{ E_{r}, E_{\phi } \}$ 
are radial and angular components 
of the field are given 
by Eq.\,(\ref{Ei3}), which is found for the plasmon 
resonance in the metal  nanoring 
(see Figure \ref{Fig2}),
$r \equiv \left| w \right| = \sqrt{u^{2} + v^{2}}$ 
is the radius in $\left\lbrace u,v \right\rbrace $ 
frame. The values of 
$u = \exp(x) \cos (y) + x_{0}$ 
and 
$v = \exp(x) \sin (y)$ 
are obtained from Eq.\,(\ref{zw}). 

\begin{figure}[ht]
	\centering{\includegraphics[scale=0.3]{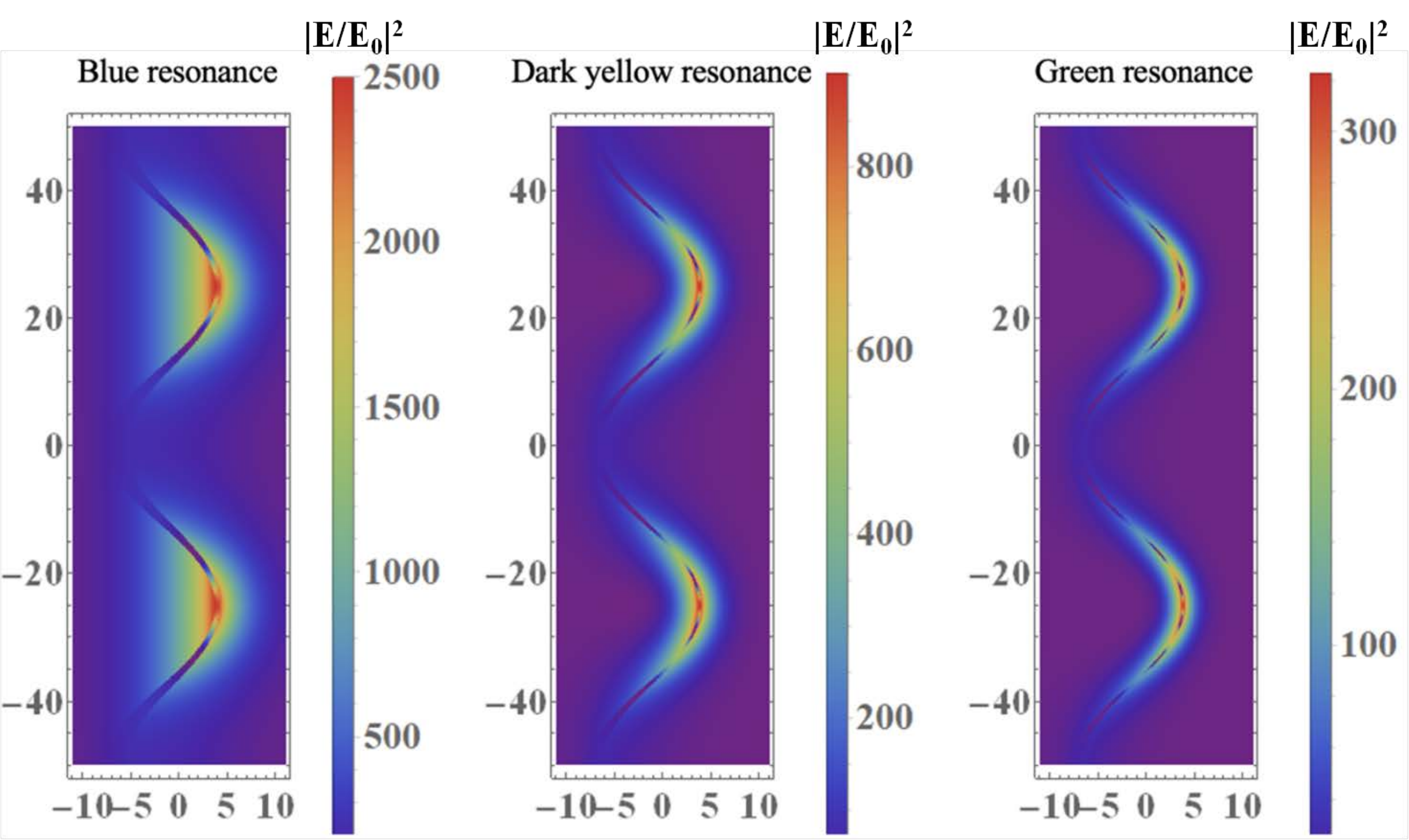} }
	\caption{Intensity of local electric field in a 
		modulated silver film for the blue, dark yellow, 
		and green resonances 
		displayed in Figure \ref{Fig3}b, where the period is 
		$L = 50 \,nm$, modulation amplitude is 
		$h = 11\, nm$, silver permittivity 
		at $\lambda = 785 \, nm$ is 
		$ \varepsilon_m = -30 + 0.4i$, permittivity 
		of dielectric substrate is 
		$\varepsilon_d =2$, and permittivity of front space 
		$\varepsilon_1 = 1$. }
	\label{Fig4}
\end{figure}

\begin{figure}[ht]
	\centering{	\includegraphics[scale=0.4]{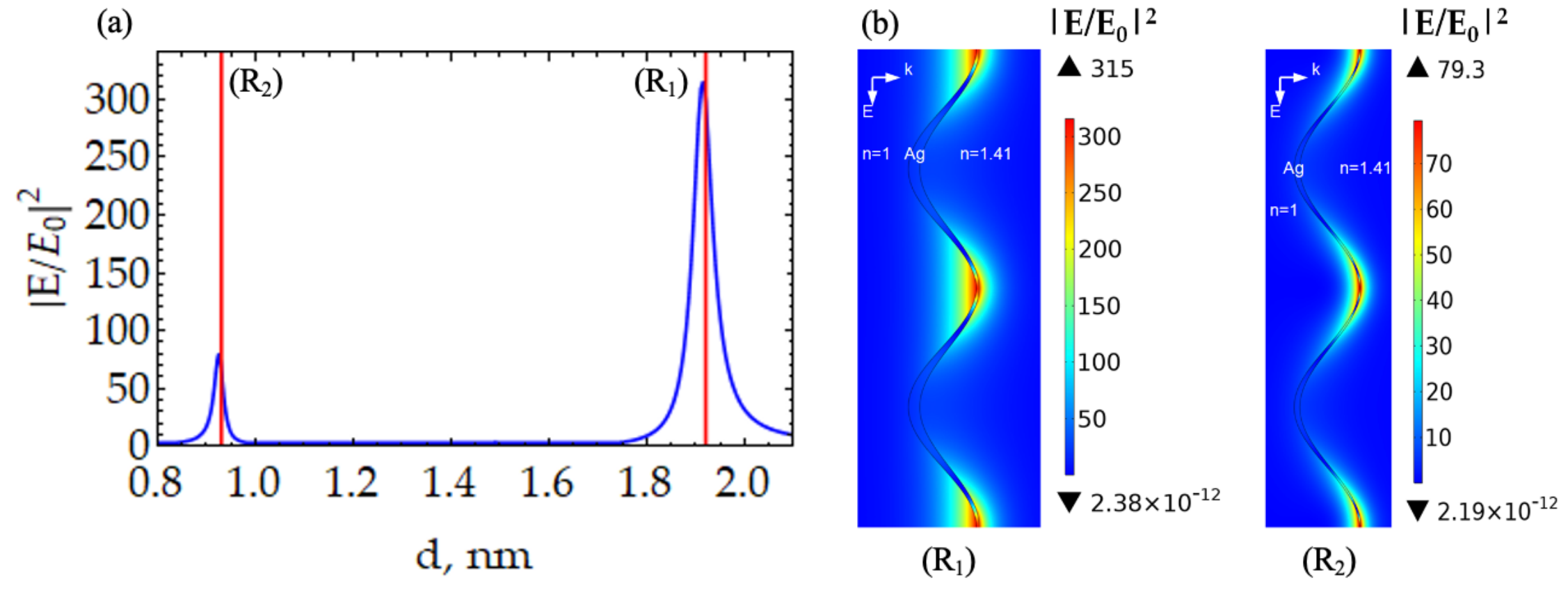}}
	\caption{(a) Electric field intensity for different 
		hicknesses 
		\textit{d} of the silver film (illumination 
		under normal incidence, with the following 
		parameters: \textit{L} = 50 \textit{nm} and
		\textit{h} = 11 \textit{nm}). 
		The red vertical lines indicate resonance thicknesses 
		obtained from analytical theory. 
		(b) Distributions of the electric field intensity 
		corresponding to two resonances ($R_1$) and 
		($R_2$) displayed in figure \ref{Fig5}a.}
	\label{Fig5}
\end{figure}

\begin{figure}[ht]
	\centering{ \includegraphics[scale=0.4]{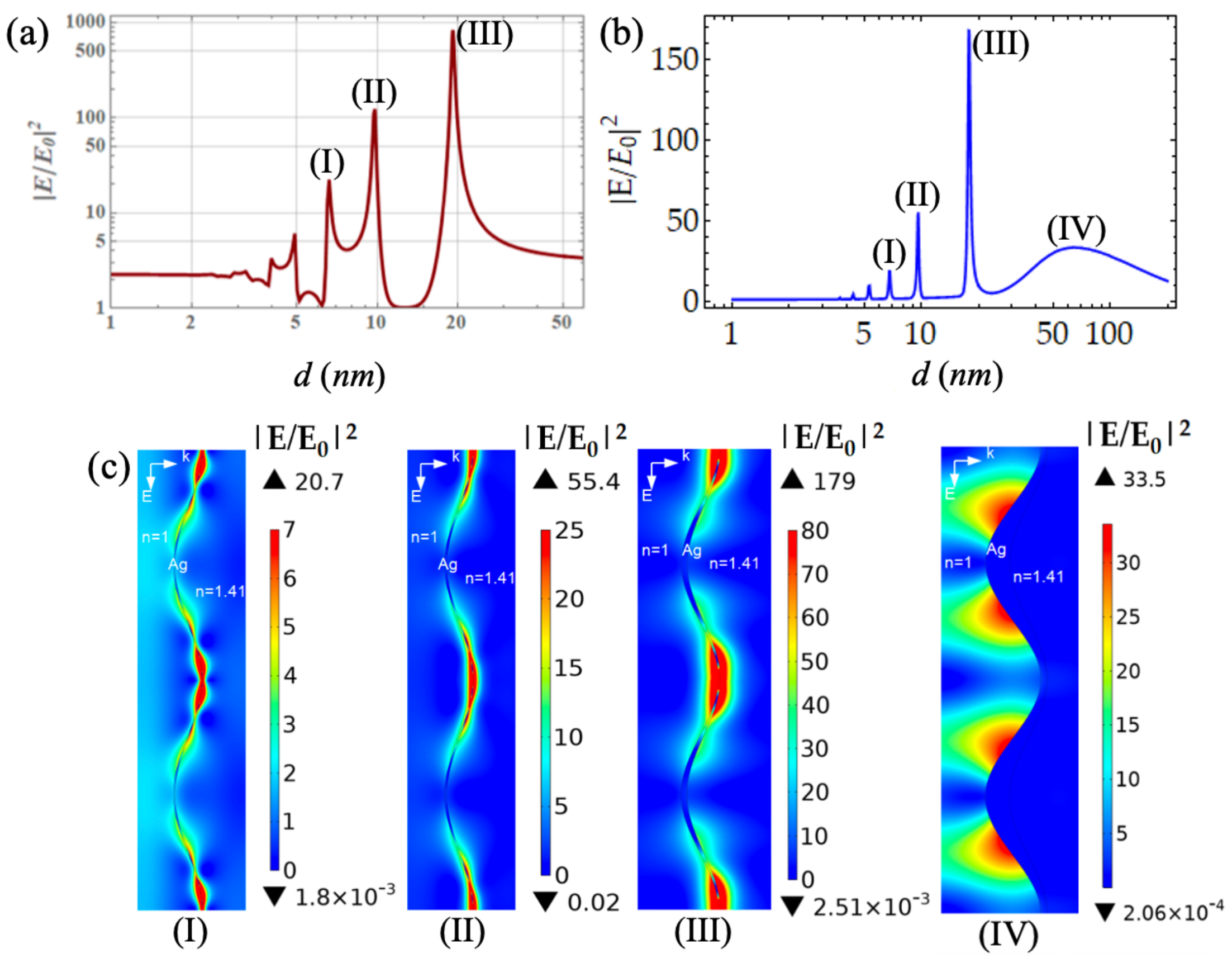}}
	\caption{
		Intensity of electric field for 
		different thicknesses of silver film \textit{d} with 
		the following parameters: period 
		\textit{L} = 750 \textit{nm} and amplitude of the modulation
		\textit{h} = 100 \textit{nm} (illumination at 
		$\lambda$ = 785 \textit{nm}, under normal incidence), 
		calculated by using
		\textbf{\textit{(a)}} the result of Eq.\,(\ref{EmaxXY}) and 
		\textbf{\textit{(b)}} computer simulations. 
		\textit{\textbf{(c)} } 
		Computer simulation of distribution 
		of local electric field corresponding to 
		resonances: (I), (II), (III) and (IV) displayed in 
		figure  Fig.\,6b.}
	\label{Fig6}
\end{figure}

Below, we suppose that $E_{0} = 1$ and measure 
all fields in terms of the applied field. The local 
electric field at the modulated metal surface in the
point with coordinates
\begin{eqnarray}
X_{max} =(L/2 \pi) x^{(f)}_{max} =
(L/2 \pi) \log(1 + x_{0}),\nonumber \\
Y_{max} = L/2 + m L \,( m = 0, \pm 1, \pm 2, \ldots) \nonumber
\end{eqnarray}
of the modulated silver film is shown in 
Figure \ref{Fig3}b as a function of the film 
thickness $d$. The wavelength of the incident light 
is $\lambda = 785\,nm$. The film is deposited on 
the dielectric with permittivity $\varepsilon_{d}$, and
the permittivity of the front medium 
$\varepsilon_{1} = 1$. 
The surface electric field is much enhanced for the 
film thickness corresponding to the excitation of 
plasmon resonances. The resonance thickness
$ d_{res,n} $ is obtained
from Eqs.\,(\ref{dRes} ) and (\ref{xd}):
\begin{equation}\label{dResXY}
	d_{res}=\frac{L}{2 \pi }
	\log   \left[ \frac{
		\left(\frac{\left(\varepsilon_1-\varepsilon_{m1}\right)
			\left(\varepsilon_d-\varepsilon_{m1}\right)}
		{\left(\varepsilon_1+\varepsilon_{m1}\right) 
			\left(\varepsilon_d+\varepsilon_{m1}\right)}
		\right)^{\frac{1}{2 n}} -\tanh \frac{ h_1}{2} }
	{1 - \tanh \frac{ h_1}{2} }\right]
\end{equation}
where $ L $ is the period  and $ h_{1} = h (2 \pi /L) $ is  the
dimensionless film modulation. 
The resonances are indicated by vertical lines in Figures
\ref{Fig3}b and \ref{Fig5}a. 
The distributions of the local electric field for 
the three resonances are displayed in 
Figure \ref{Fig4}. 
The local electric field is much enhanced and 
exceeds the incident field $E_{0}$ by two-three 
orders of magnitude. The enhanced field spreads 
over the grating achieving its maxima in the 
depressions of the front surface. 
The amplitude of the maximum field 
in the point with coordinates 
$ X_{max}, Y_{max} $
at the front surface of the metal film 
(star $" \ast "$ in Fig.\,\ref{Fig3}a) is given
by the following equation obtained from
Eq.\,(\ref{Ei3}) by substitution there
$ r = 1 $ and $ f = \pi $ and multiplying the result by 
matrix (\ref{MM}) 
\begin{widetext}
\begin{eqnarray}
&	E_{x} (X_{max}, Y_{max}) = 
E_{0} (1 - x_{0}) \left[ \frac{1}{1 + x_{0}} + \sum_{n=1}^{\infty} \left(-x_0\right)^n 
\frac{\left(\varepsilon_1+
\varepsilon_m\right) 
\left(\varepsilon_m -\varepsilon_d\right)+
\left(d_0+1\right)^{2 n} 
\left(\varepsilon_1-\varepsilon_m \right)
\left(\varepsilon_d+
\varepsilon_m \right)}
{\left(\varepsilon_1-\varepsilon_m \right)
\left(\varepsilon_m -\varepsilon_d\right)+
\left(d_0+1\right){}^{2 n}
\left(\varepsilon_1+\varepsilon_m \right) 
\left(\varepsilon_d+\varepsilon_m\right)}
\right], \nonumber \\
&	 E_{y} (X_{max}, Y_{max}) = 0
	\label{EmaxXY} 
\end{eqnarray}
\end{widetext}

where the parameters $ x_{0} $ and $ d_{0} $
are given by Eqs.\,(\ref{xd}). 
The important characteristic is the amplitude of the
resonance field averaged over the
surface of the metal film, which has resonance thickness
$ d_{res,n} $
given by Eq.\,(\ref{dResXY}). 
The amplitude is
still given by Eq.\,(\ref{Imax}), where the
ratio $ x_{0}/r_{0} $ is replaced by
$ \tanh \frac{ h_1}{2} $. 
For the practically important case 
$ \left| \varepsilon_{m} \right| \gg \varepsilon_{d}$,
the averaged intensity at the 
$n$-$th$ resonance is estimated as follows:
\begin{equation} \label{ImaxXY} 
	I_{\max ,n} \sim
	\left[ 
	\frac{\epsilon_1 \epsilon _{m1}}
	{ \epsilon_{m2} 
		\left(\epsilon_d+\epsilon_1\right)}
	\tanh^{ n} \frac{ h_1}{2} 
	\right]^{2} 
\end{equation}
where $ h_{1} = h (\frac{ 2 \pi }{ L } )$ and
we skip numerical coefficient on the order one. 
The analytical results are compared with results
of computer simulations. Computer 
simulations are performed for 
the electromagnetic field distribution in 
a weakly profiled silver film deposited on a 
photoresist. The dimensions of 
the computer model are the same than those described in 
Figures \ref{Fig3}b and \ref{Fig4} 
($ \varepsilon_{1} = 1 $, $ \varepsilon_{d} = 2 $,
i.e., the refractive index of the photoresist is $n=1. 41$). 
The simulations are done in COMSOL
environment. The Maxwell equations are 
solved by the finite element method (FEM). 
We demonstrate the enhancement of the electric 
field $|E/E_{0}|^{2}$ in the modulated
thin metal film with period
$L=50\;nm $ 
and amplitude of the modulation
$ h=11\;nm $ (see Figure \ref{Fig5}). 
It can be seen that positions of the analytical and 
numerical resonances are in a good agreement. 
The distribution of the electric field for two 
resonances shown in Figure \ref{Fig5}a are 
displayed in Figure \ref{Fig5}b. Computer 
simulations are 
in agreement with analytical theory
(see the blue and dark yellow 
resonances in figure \ref{Fig4}). We 
also simulate the resonant enhancement of the 
electric field for a modulated metal film 
whose parameters correspond to the 
experiment discussed below (see Figure 
\ref{Fig6}). 
Figure \ref{Fig6}a,b show the 
distribution of the electric field intensity calculated
with the result of Eq.\,(\ref{EmaxXY}) of the analytical theory,
and computer simulations, respectively. 
Resonances are due to the excitation of 
dipolar and multipolar plasmons 
in the modulated metal film. 
The electric field distribution in dipolar 
and multipolar plasmons is in qualitative agreement with 
analytical theory [see the resonances: (I), (II), and (III) 
displayed in figure \ref{Fig6}a,b]. 
The field distribution in the resonance
(IV) in Figure \ref{Fig6}b,c differs from all other resonances. 
The electric field expands from the metal
surface on the distance comparable with the wavelength
and period of the modulation. We speculate that
in this case the plasmon 
resonance hybridizes with plasmon propagating over
the modulated metal surface (see, e.g.,
\cite{DykhneSarychev2003}). The hybridization 
could be the subject for the further consideration. 
In summary, we present the analytical theory
of plasmons excited in a modulated metal film 
deposited on a dielectric substrate. 
Computer simulations and results given 
by the analytical theory are in a good agreement. 
The resonance conditions are found as a function 
of the metal permittivity, 
the modulation and the thickness of the film. 
When the grating is excited by the
incident light, the surface electric field is much 
enhanced. The quasistatic theory qualitatively 
described the field enhancement even 
in the case $k L \simeq 1$, 
since the field is concentrated in the 
nanograting whose thickness and modulation are 
much smaller than $\lambda$. The size of a plasmon maxima $l_{p}$ is on the order of the film modulation $l_{p} \approx h$ (see Figs.\ref{Fig4},\ref{Fig5},\ref{Fig6}). The radiation loss, which is propotional to $l_{p} d k^{2}$ remains to be less than one. For example, for silver film with parameters $h=100 \,nm, d=20 \, nm$ and $\lambda=785 \, nm$ (see Fig.\ref{Fig6}), $l_{p} d k^{2} \approx 0.1<1$.
The positions of the plasmon resonances from computer 
simulations are almost coincide with results of 
Eq.\,(\ref{dResXY}). The resonance enhancement of the 
electric field obtained in computer simulations is very large. 
Yet, the computer results are few times smaller than results 
of the quasistatic Eq.\,(\ref{EmaxXY}). 
The discrepancy is due to the radiation loss that
can be taken into account by using a perturbation theory. 
Yet, we do not expect qualitative difference between
the quasistatic approximation and exact result 
while plasmons are localized like in resonances (I) - (III)
(Figure \ref{Fig6}). 
The developed analytical theory can be used for the
design of new SERS substrates and other optical sensors. 
The theory can be extended to multilayered 
metal-dielectric films. 
\section{Experimental investigation of holographic metasurfaces}
\subsection{Fabrication process}
\begin{figure}[ht]
	\centering{	\includegraphics[scale=0.3]{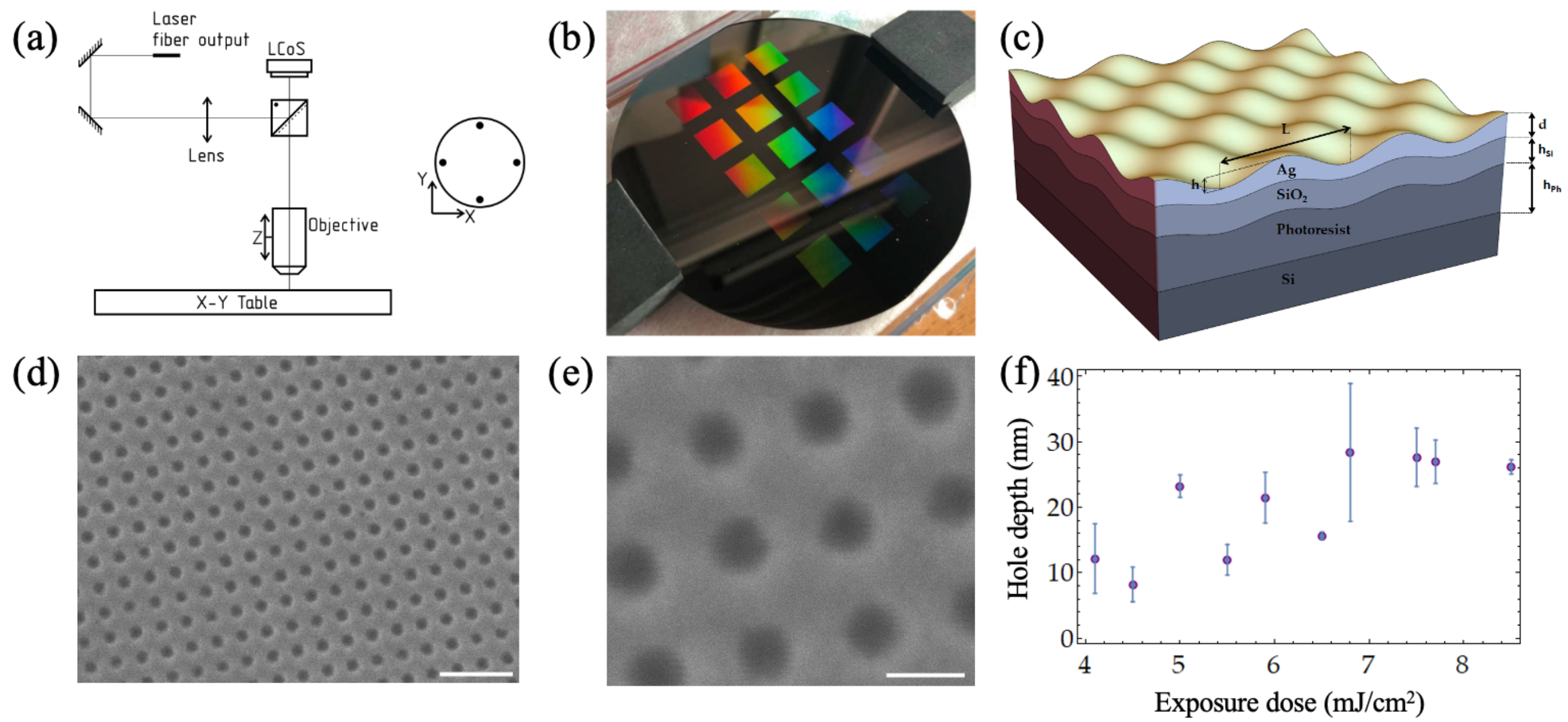}}
	\caption{(a) Principle scheme of the recording 
		setup. (b) Photo of 4-inch wafer where holographic 
		metasurfaces are fabricated. (c) Principle scheme 
		of the metasurface design with the following 
		parameters: period is $L = 720-770$ $nm$, 
		modulation amplitude is $h = 8-30$ $nm$, silver 
		layer thickness is $d = 60$ $nm$, thickness of 
		$SiO_2$ is $h_{Si}$ = 80 $nm$, photoresist 
		thickness is $h_{Ph} = 1. 5-1. 9$ $\mu m$. SEM 
		images of the metasurfaces (Period $L \approx$ 
		760 $nm$, and film modulation $h \approx$ 30 
		$nm$) with a scale bar of 2 $\mu m$ for (d) and 
		500 $nm$ for (e). (f) Hole depth versus exposure 
		dose (depth measured by AFM).}
	\label{Fig7}
\end{figure}

The metasurfaces are fabricated by the following 
steps. As a first step $1. 5-1. 9\;\mu m$ recording 
layer of positive photoresist (Microposit S1800) is 
deposited by spin-coating on the silicon wafer. 
Then, fifteen areas of two-dimensional modulated 
films ($1. 5 \times 1. 5\;cm^{2}$) are formed by 
holographic lithography technique. The principle 
scheme of recording setup employs a 
phase-modulator HED-6001 Monochrome LCOS, 
and an ultraviolet laser ($\lambda=405\;nm$) 
(see Figure \ref{Fig7}a). 
Exposure dose is varied 
from 4. 1 to 8. 5 $\mu J/cm^{2}$. The angle between 
interfering light waves is around 49$^{\circ}$. Next, 
the resist is developed in Microposit 303A 
developer. Afterwards, two successive depositions 
of a 80-\textit{nm} $SiO_{2}$ layer and a 
60-\textit{nm} silver layer are realized by electron 
beam evaporation. The metasurfaces with period 
from $720\,nm$ to $770\,nm$ and modulation 
amplitude from 8 \textit{nm} to 30 \textit{nm} are 
fabricated (see Figure \ref{Fig7}b). 
The principle 
scheme of the metasurface morphology is shown in 
Figure \ref{Fig7}c. 
The morphology of the 
fabricated metasurfaces is measured with a scanning 
electron microscopy (SEM, JEOL), and SEM images 
of these metasurfaces are shown in Figure 
\ref{Fig7}d,e. 
Finally, modulation amplitude 
measurements are made with an atomic force 
microscopy (AFM; Solver Pro NT-MDT) for 
different exposure doses (see Figure 
\ref{Fig7}f). 
\subsection{Deposition of 4-Mercaptophenylboronic%
	Acid for SERS }
In order to study the sensitivity of the holographic 
metasurfaces, we employed molecules of 
4-mercaptophenylboronic acid (4-mPBA), which 
are small molecules (thickness of a 4-mPBA 
monolayer is around 0.8 $nm$ \cite{Barriet2007}). 
We have chosen the boronic acid, which is a 
specific substance because the boron group 
covalently binds diols and forms a boron ester. 
Therefore, boronic acid is considered as a specific 
agent for sugars and their fragments. There is a 
strong covalent fixation of this acid on the surface 
of silver due to the thiol group (–SH). The analyzed 
molecules are covalently bound to the silver, 
thereby increasing the probability of the location of 
molecules in the region of generation of giant 
electromagnetic fields. In addition, we prepared 
solutions of 4-mPBA in ethanol with concentrations 
varying from 2. 3 $\times$ 10$^{-8}$ M to 2. 3 $\times$ 10$^{-3}$ M. 
Then, for~SERS experiments, an aliquot of each 4-mPBA 
solution was applied onto holographic metasurface 
and then dried at room temperature. 
\subsection{Optical Characterization}
For characterizing the optical properties of 
metasurfaces, the reflectance spectra have been 
recorded by using a spectroscopic ellipsometer SE 
850 DUV (Sentech, Germany). For~the SERS 
measurements, a Raman spectrometer InnoRam 
(BWTek, USA) with an excitation wavelength of 
785 $nm$ and a spectral resolution of 4 $cm^{-1}$ 
was employed. The~acquisition time and the laser 
power have been set at 1 s and 3. 84 mW, 
respectively. A~microscope objective ($\times$50, 
NA = 0.8) was used in order to focus the laser beam 
on the metasurface, and~the Raman signal of the 
4-mPBA molecules immobilized on metasurface 
was detected by this same objective in a 
backscattering configuration. 
\section{Results and discussion}
\subsection{Optical properties of holographic%
	metasurfaces}
\begin{figure}[ht]
	\centering{	\includegraphics[scale=0.3]{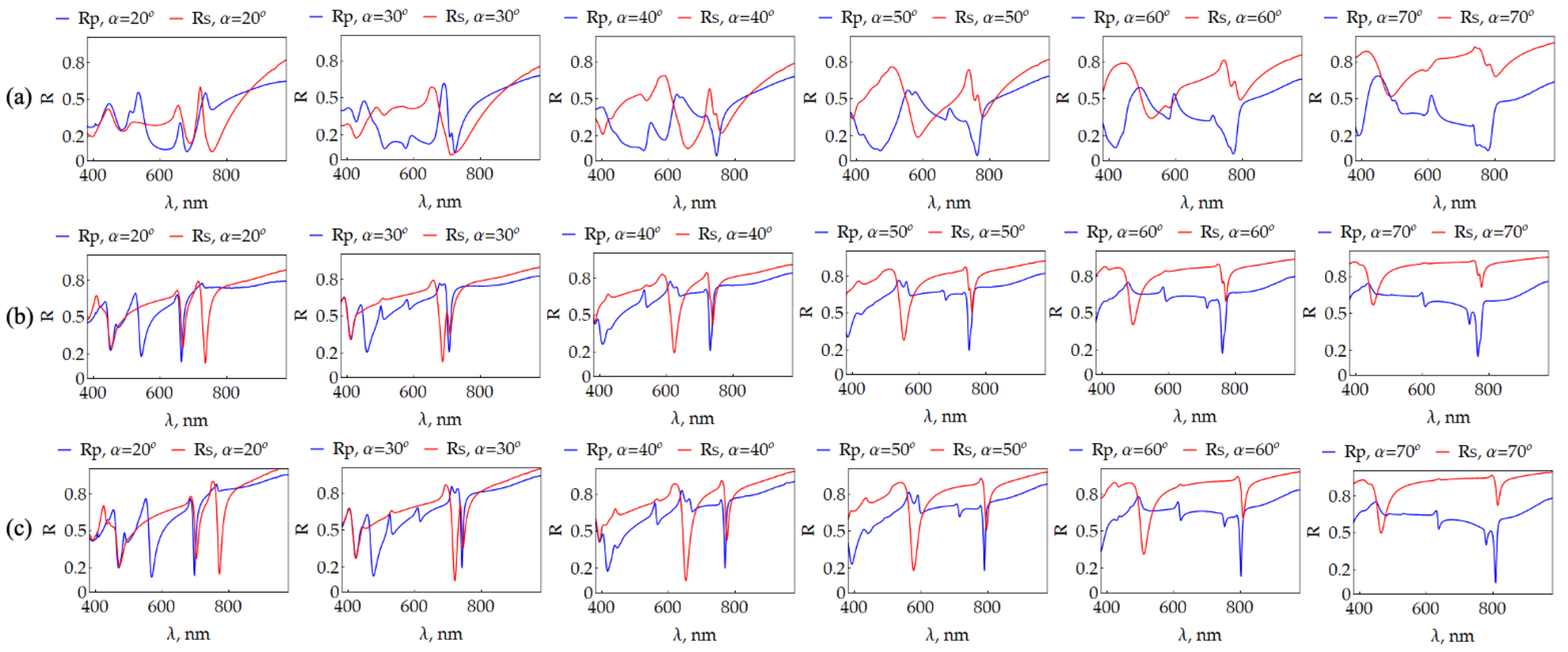}}
	\caption{Experimental angular reflectance spectra 
		from the metasurface for different angles of the 
		incident light and with the following structure 
		parameters: (a) $L$ = 720 $nm$, depth $h$ 
		$\approx$ 30 $nm$; (b) $L$ = 720 $nm$, depth 
		$h$ $\approx$ 20 $nm$; (c) $L$ = 770 $nm$, 
		depth $h$ $\approx$ 18 $nm$. }
	\label{Fig8}
\end{figure}

\begin{figure}[ht]
	\centering{	\includegraphics[scale=0.4]{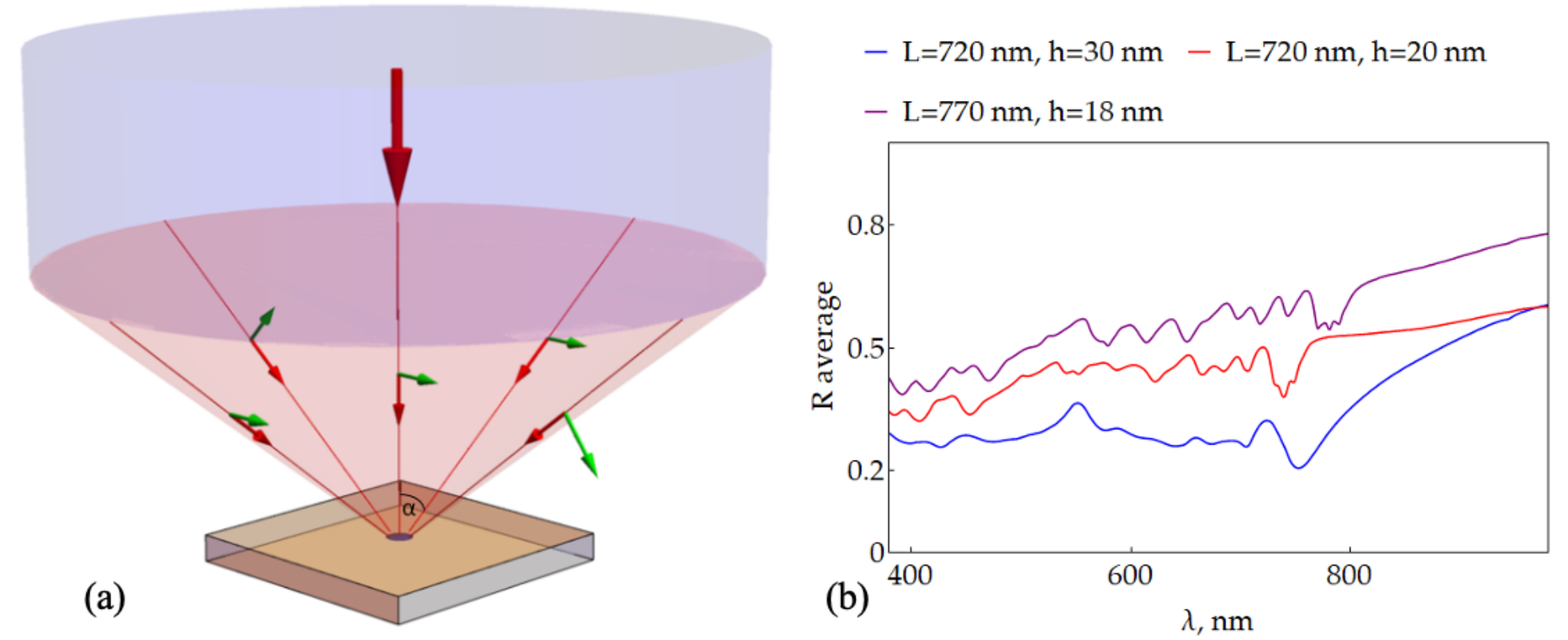}}
	\caption{(a) 
		Schematic illustration of the polarized 
		light impinged to the sample by the lens. The red 
		and green arrows show the directions of the 
		wave-vector and the electric field, respectively. 
		(b) Average experimental reflectance spectra 
		from modulated silver film (metasurface), the 
		reflectance is averaged over the light cone, which 
		is focused by the Raman microscope on the 
		sample.}
	\label{Fig9}
\end{figure}

To determine the macroscopic optical properties of 
metasurfaces, the angular reflectance spectrum is 
measured by ellipsometry in the range from 
$20^{\circ}$ to $70^{\circ}$. The 
reflectance spectrum is measured for two 
polarizations: $Rp$ for $p$-polarization of the 
incident light and $Rs$ for $s$-polarization of the 
incident light. The measured reflectance spectra for 
$Rp$ and $Rs$ show deep gaps with a complex 
behavior versus period and angle 
(see Figure \ref{Fig8}). 
Raman instruments are usually attached to 
microscopes, where the excitation and the 
inelastically scattered light are delivered. The light is 
collected from the sample through the same 
objective in a backscattering geometry. The 
objective is characterized by a numerical aperture 
(NA) equal to 0.8 in our case. This means that the 
cone angle is equal to 
$2 \alpha=\arcsin{(NA)}=106^{\circ}$. So, we have 
to take into account all the rays of the incident light 
at angles $\alpha$ 
from $0^{\circ}$ to $53^{\circ}$ degrees. The focal 
length $f$ is related to the lens diameter by the ratio 
$(D/2f)=\tan\alpha$ ($ D $ - lens diameter). We find 
the average reflection in the approximation of the 
geometric optics by integrating the reflection over 
the basis of the cone of the incidence. The principle 
scheme of the polarized light collected by the lens is 
displayed in Figure \ref{Fig9}a. The average 
reflection is expressed as follows:
\begin{equation}\label{Rav}
	R_{average}=2(\frac{1}{NA^2}-1) 
	\int_{0}^{\alpha_{max}} \frac{\sin \alpha}{\cos 
		\alpha^3} R(\alpha) d\alpha 
\end{equation}
where $\alpha_{max}=\arcsin{(NA)}$. In addition, 
each ray has a transverse and longitudinal field 
component, so we average $Rp$ and $Rs$. The 
resulting average reflection is displayed in Figure 
\ref{Fig9}b. 
We observe minima on the spectral range from 
$400\;nm$ to $800\;nm$ and more specifically close 
to the excitation wavelength of $785\;nm$. We 
assume that these minima in the 
reflectance spectrum correspond to plasmon 
resonances. These plasmonic resonances (close to 
the excitation wavelength of $785\;nm$) will permit 
the enhancement the Raman signal of 4-mPBA 
molecules. 
\subsection{SERS detection of 4-mercaptophenylboronic acid molecules}
\begin{figure}[ht]
	\centering{\includegraphics[scale=0.45]{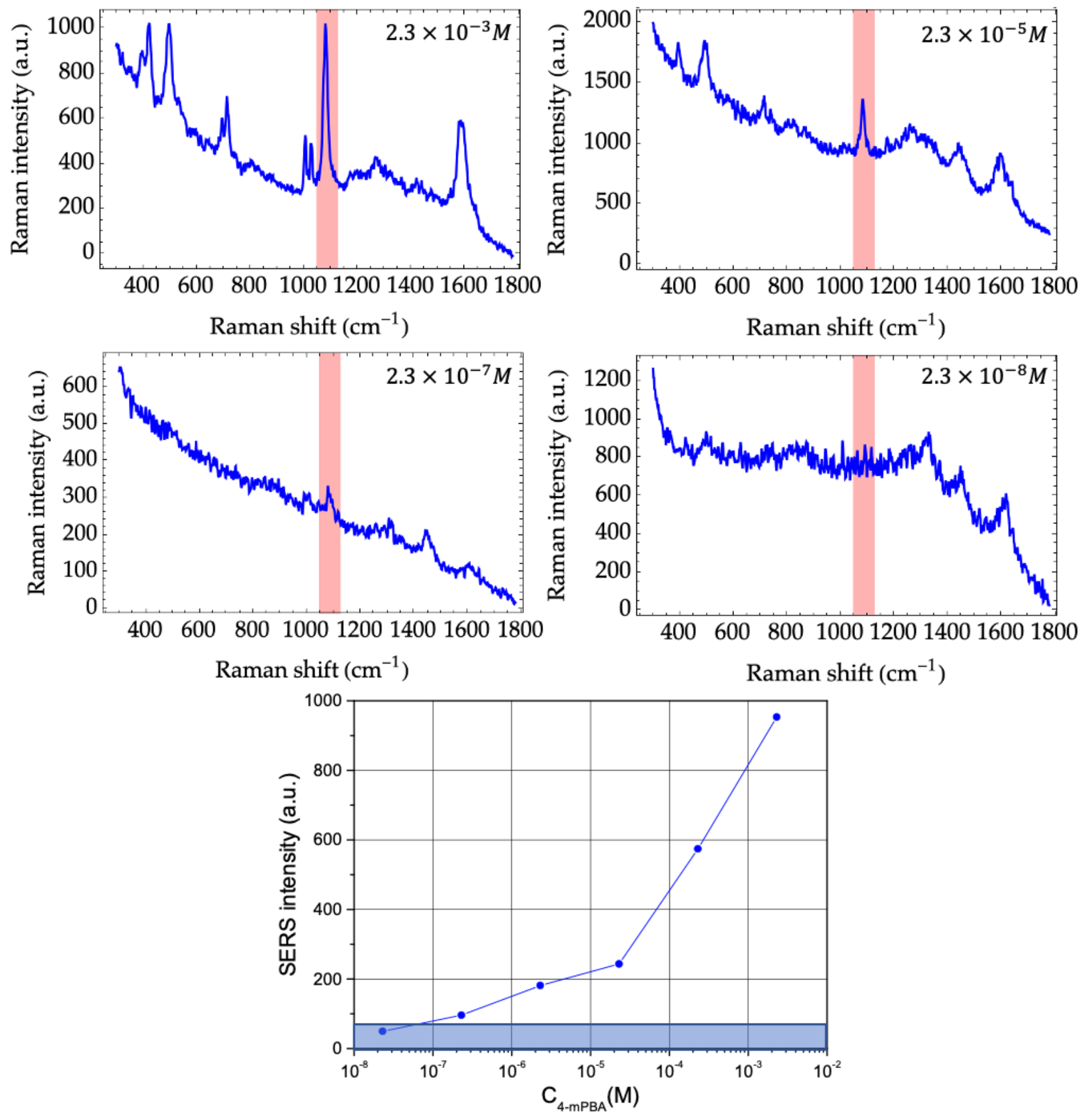}}
	\caption{ SERS
		spectra of 4-mPBA molecules 
		recorded on the metasurface for the 
		concentrations of 2. 3 
		$\times$ 10$^{-3}$ M, 2. 3 $\times$ 10$^{-5}$ M, 
		2. 3 $\times$ 10$^{-7}$ M and 2. 3 $\times$ 10$^{-8}$ M. 
		At the bottom of the figure, graph of SERS intensity 
		versus 4-mPBA concentration, where the blue zone 
		corresponds to the zone of the 3 $\times$ noise level. }
	\label{Fig10}
\end{figure}

For all the SERS experiments, a 
modulated silver metasurface with a period 
$L=760 \; nm, h=30 \; nm$ is used. 
This metasurface has the maximum available 
modulation depth $h$ = 30 $nm$. With a higher 
modulation depth, the number of electromagnetic 
modes increases. Thus, this modulation allows 
obtaining deep minima in the reflectance spectrum 
close to the excitation wavelength of $785\;nm$ (see 
figure \ref{Fig9}b). Then, to evaluate the sensitivity 
of these metasurfaces, 4-mPBA molecules are 
grafted onto them by using the functionalization 
protocol described in section 3.2. Next, the SERS 
spectra have been recorded at the excitation 
wavelength of 785 $nm$. Figure \ref{Fig10} shows 
the SERS spectra of 4-mPBA molecules on 
metasurfaces obtained for each concentration 
varying from 2. 3 
$\times$ 10$^{-8}$ M to 2. 3 $\times$ 10$^{-3}$ M. 
From the SERS spectrum obtained for the concentration 
of 2. 3 mM depicted in Figure \ref{Fig10}, seven 
characteristic Raman peaks of 4-mPBA molecules 
are well-distinguished 
\cite{Nechaeva2020,SuH2017,SunF2014}. The 
Raman peaks at 420 $cm^{-1}$, 693 $cm^{-1}$ and 
1072 $cm^{-1}$ correspond to the C--C--C in-plane 
bending mode coupled with C--S stretching mode 
(called respectively: $\beta_{CCC}$ and 
$\nu_{CS}$). Then, the peak at 480 $cm^{-1}$ 
corresponds to the C--C--C out-of-plane bending 
mode coupled to O--B--O in-plane bending mode 
(called respectively: $\gamma_{CCC}$ and 
$\beta_{OBO}$). The Raman peaks at 1000 
$cm^{-1}$ and 1030 $cm^{-1}$ correspond, 
respectively, to the C--C--C in-plane bending mode 
(called $\beta_{CCC}$) and the C--H in-plane 
bending mode (called $\beta_{CH}$). Finally, the 
Raman peak at $1580\; cm^{-1}$ corresponds to the 
C--C stretching vibrational modes (called 
$\nu_{CC}$). For determining the detection limit, 
we have chosen the most intense Raman peak, 
which is located at 1072 $cm^{-1}$, and a detection 
limit of 230 nM was achieved for the sensing of 
4-mPBA molecules (see Figure \ref{Fig10} where 
the blue zone corresponds to the zone of 
the 3 $\times$ noise level for the last curve) with
a signal-to-noise ratio $> 3$. 

\begin{figure}[ht]
	\centering{	\includegraphics[scale=0.5]{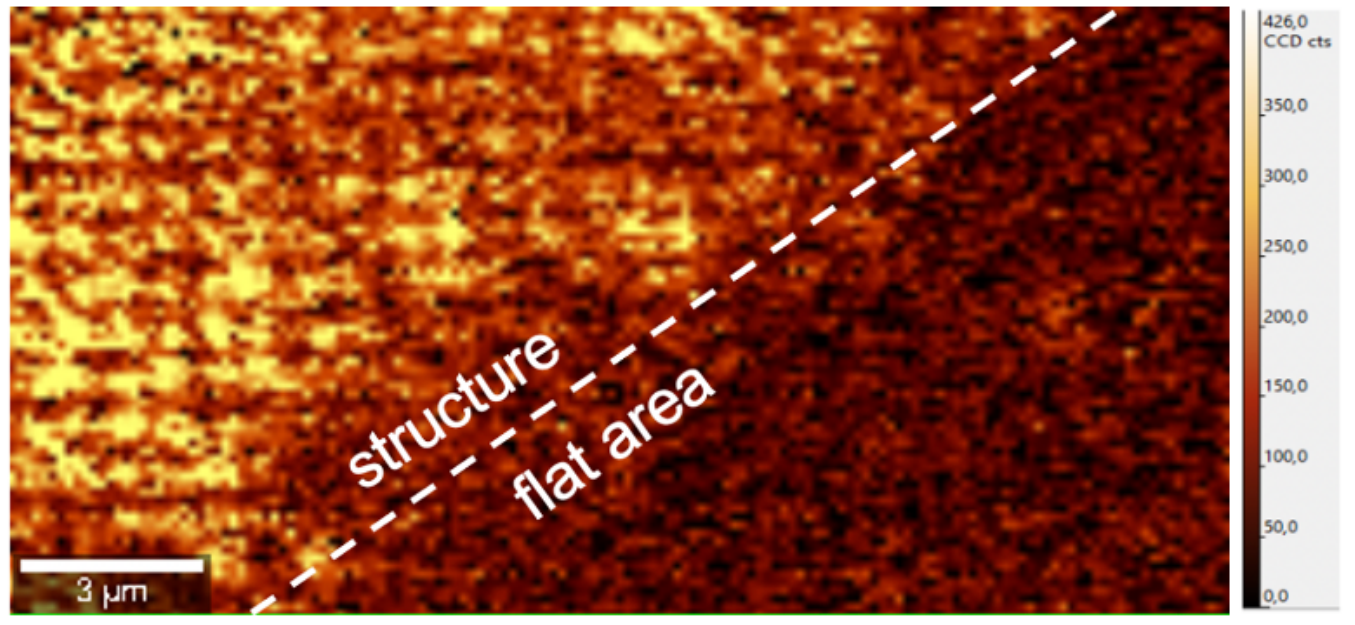}}
	\caption{Optical image of the intensity distribution 
		for the Raman bands 1060--1100 	$cm^{-1}$. }
	\label{Fig11}
\end{figure}

In addition, the distribution of Raman 
hotspots generated by 
periodic holes (grid) repeats the morphology of the 
surface, which is shown on the optical image 
displayed in Figure \ref{Fig7}. 
Therefore, we can see the SERS effect distributed 
uniformly over the entire metasurface. 
The pattern of the Raman
signal in Figure \ref{Fig11} repeats the periodic structure
of the modulated film. Therefore, the maxima of the Raman
signal repeat the maxima of the plasmon field $ I(x, y) $. 
One could suggest that the SERS is a local effect, namely,
$ G (x, y) \propto I(x, y)^{2} $. 
Investigation of the modulated metal film 
with progressively smaller period $ L $
can confirm locality of SERS effect. 
\section{Conclusion}
In this paper, we presented the low-cost fabrication 
and the study of a resonant metasurface consisting 
of a periodically modulated metal film that exhibits 
an anomalous optical response due to the excitation 
of localized surface plasmons. The distribution of 
the enhanced electric field generated by the film 
modulation has repeated the surface morphology, 
and this distribution was uniform over this entire 
surface. The holographic metasurface allowed 
achieving a detection limit of 230 nM for the 
4-mPBA molecules serving here as a
proof-of-concept of SERS sensing. Thus, the 
low-cost and easy-made holographic SERS 
metasurfaces modified with 4-mPBA molecules can 
be used for quantitative biosensing of glycated 
proteins and sugars. Moreover, the analytical theory 
can allow the design and the optimization of 
various plasmonic sensors. 
\section{Funding}
This work was supported by the Russian 
Foundation for Basic Research (Grant No. 
20-21-00080). The Raman measurments was 
supported by the Russian Science Foundation 
(Grant No. 21-79-30048).

\end{document}